\documentclass[conference]{IEEEtran}
\IEEEoverridecommandlockouts
\usepackage{cite}
\usepackage{amsmath,amssymb,amsfonts}
\usepackage{amsthm}
\usepackage{booktabs}
\usepackage{multirow}
\usepackage{algorithmic}
\usepackage{graphicx}
\usepackage{textcomp}
\usepackage{xcolor}
\def\BibTeX{{\rm B\kern-.05em{\sc i\kern-.025em b}\kern-.08em
    T\kern-.1667em\lower.7ex\hbox{E}\kern-.125emX}}
\begin{document}

\title{
% Weakly Supervised Indoor Relative Localization from WiFi Fingerprint Traces
% Learning Displacement-Aware WiFi Representations for Weakly Supervised Relative Localization
Learning Displacement-Aware WiFi Representations for Weakly Supervised Relative Localization}

\author{
    \IEEEauthorblockN{
        Tzu-Ti Wei, 
        Po-Cheng Chen, 
        Yu-Chee Tseng, 
        Jen-Jee Chen
    }
    \IEEEauthorblockA{
        \textit{College of AI, National Yang Ming Chiao Tung University, Taiwan}\\
        \{
            a2699560.ai09, 
            havefree456.c,
            yctseng, 
            jenjee
        \}@nycu.edu.tw
    }
}
\maketitle

\begin{abstract}
% WiFi fingerprint-based indoor localization has been intensively studied, but most existing methods target absolute localization and require dense coordinate annotations, which are costly to obtain at scale. In this paper, we study a different problem, namely indoor relative localization: given two WiFi fingerprint traces, estimate the displacement between their endpoints. To reduce annotation burden, we use weak supervision in the form of stepwise motion vectors obtained from inertial sensing, instead of strong absolute labels. Based on paired fingerprint traces (f-traces) and displacement traces (d-traces), we develop a cross-modal learning framework called \emph{Intersection Pathway (IP)}. IP aligns the two modalities in a shared latent space, where latent addition and subtraction preserve displacement composition for direct relative-displacement inference. Experiments on a near-real indoor dataset show that the proposed design learns displacement-aware WiFi representations and yields strong relative-localization performance. The learned model can also be extended to few-shot absolute localization with sparse anchors, achieving sub-meter accuracy while requiring far fewer strong labels.

WiFi fingerprint-based indoor localization has been widely studied, but most existing approaches focus on absolute positioning and rely on dense coordinate annotations, which are costly to obtain at scale. In this paper, we study a fundamentally different problem: \emph{relative localization}, where the goal is to directly estimate the displacement between two WiFi fingerprint traces without predicting their absolute positions. To reduce annotation overhead, we adopt weak supervision in the form of stepwise motion vectors obtained from inertial sensing. We propose \emph{Intersection Pathway (IP)}, a cross-modal learning framework that aligns \emph{fingerprint traces (f-traces)} and \emph{displacement traces (d-traces)} in a shared latent space. The key idea is to enforce an \emph{additive structure} in the latent space, such that latent addition and subtraction correspond to physical motion composition, enabling direct relative-displacement inference. Experiments on a synthesized dataset derived from real measurements demonstrate that the proposed method learns \emph{displacement-aware} WiFi representations and achieves accurate relative localization across varying displacement ranges. Furthermore, the learned model can be extended to few-shot absolute localization with sparse anchors.


\end{abstract}
\section{Introduction}
\label{ch:intro}

% Indoor positioning is a key enabler of location-based services, robotics, and smart-space applications \cite{liu2020survey, pasricha2020overview}. Existing indoor-localization techniques include RF-, vision-, inertial-, and UWB-based methods \cite{4343996, zhou2024vtil, harle2013survey, grosswindhager2019snaploc}. WiFi fingerprinting is especially attractive because WiFi infrastructures are already widely deployed \cite{shang2022overview}. Typical WiFi localization systems map RSSI/CSI fingerprints, or short sequences of fingerprints, to physical positions \cite{du2020kf, localization-kuo, scrambling-kuo}, but they usually require \emph{strong} labels, i.e., location annotations of fingerprints, which are expensive to collect over a large field.

Indoor positioning is a key enabler of location-based services \cite{liu2020survey, pasricha2020overview}. Existing approaches span inertial-, RF- and vision-based techniques \cite{4343996, zhou2024vtil, harle2013survey, grosswindhager2019snaploc}. Among them, WiFi fingerprinting is particularly attractive due to the ubiquitous deployment of WiFi infrastructures \cite{shang2022overview}. Conventional WiFi localization methods map RSSI/CSI fingerprints, or short fingerprint sequences, to absolute coordinates \cite{du2020kf, localization-kuo, scrambling-kuo}. However, such approaches rely on \emph{strong} supervision—dense location annotations—which are costly and difficult to obtain at scale.

% This paper studies a different task: \emph{relative localization}. Given the WiFi fingerprint traces of two objects $a$ and $b$, we estimate the displacement from $a$ to $b$ directly, instead of first predicting two absolute positions and then subtracting them. This formulation is useful in robot following, warehouse tracking, and person-object interaction, where relative positions are more actionable than global coordinates. Our goal is not to replace app-level relative reasoning from absolute coordinates, but to ask whether relative displacement can be learned directly when dense absolute coordinate labels are unavailable. Fig.~\ref{fig:problem_definition} contrasts this setting with conventional absolute localization.

\begin{figure}[t]
    \centering
    \includegraphics[width=0.4\textwidth]{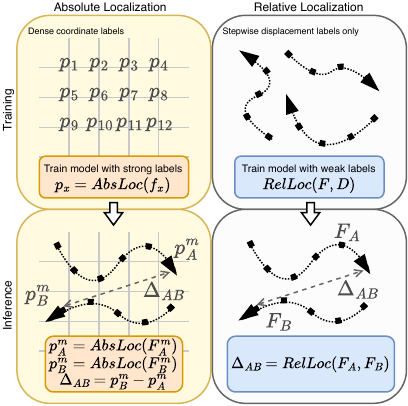}
    \caption{A conceptual comparison between conventional absolute localization and the proposed relative localization. Here, $F$ denotes a WiFi fingerprint trace and $D$ denotes a displacement trace used as weak supervision.}
    \label{fig:problem_definition}
\end{figure}

In this paper, we address a fundamentally different problem: \emph{relative localization}. Given two WiFi fingerprint traces corresponding to objects $A$ and $B$, our goal is to directly estimate the displacement $\Delta_{AB}$ from $A$ to $B$, without explicitly predicting their absolute positions. This formulation is particularly relevant in scenarios such as robot following, warehouse coordination, and human-object interaction, where relative spatial relationships are more actionable than global coordinates. This raises a key question: \emph{Can relative displacement be learned directly from weak supervision, without dense coordinate labels?} Fig.~\ref{fig:problem_definition} illustrates the conceptual difference between the proposed setting and conventional absolute localization.

Prior work on WiFi-based indoor localization has primarily focused on improving absolute positioning accuracy. Deep learning approaches have been widely adopted to mitigate fingerprint ambiguity and noise \cite{alhomayani2020deep, boundary, wang2022research, ma2023few, 9918116}, while crowdsourcing techniques aim to construct or refine large-scale fingerprint maps \cite{lashkari2018crowdsourcing, wu2017automatic, zhao2020graphips, piloc}. Systems such as BlindNavi \cite{chen2015blindnessavi} and UbiFin \cite{9451560} demonstrate the feasibility of real-world deployment. 

However, these approaches consistently rely on absolute localization with dense coordinate annotations, and do not address the problem of learning relative displacement under weak supervision. Moreover, unlike systems based on UWB, BLE tags, or WiFi FTM, our setting does not assume specialized hardware or explicit ranging capabilities, making it more broadly applicable in commodity WiFi environments.

To reduce labeling cost, we rely solely on \emph{weak} supervision in the form of stepwise motion vectors between adjacent samples, which can be obtained with low overhead from inertial sensors on robots or simple IoT devices. Our goal is to eliminate the need for dense coordinate annotations. 

To this end, we propose \emph{Intersection Pathway (IP)}, a cross-modal learning framework that makes WiFi fingerprints \emph{computable in the displacement domain}. Specifically, IP maps fingerprint traces and displacement traces into a shared latent space, where local motion relations regularize fingerprint representations. By enforcing that latent addition and subtraction correspond to physical motion composition, the model enables direct \emph{displacement-aware inference} from fingerprint traces.

% Multimodal learning has shown that heterogeneous modalities can be aligned in a shared latent space \cite{multi-modal-survey, clip-2021, girdhar2023imagebind}. Our goal is to bring these ideas together for weakly supervised relative localization.

The main contributions of this work are as follows:
\begin{itemize}
\item We formulate WiFi-based \emph{relative localization} as a weakly supervised problem, enabling direct displacement estimation without dense coordinate labels.
\item We propose \emph{Intersection Pathway}, a cross-modal framework that aligns fingerprint and displacement traces in a shared latent space, where latent addition and subtraction correspond to physical motion. This effectively induces a displacement-aware latent space in which WiFi fingerprints inherit the additive structure of physical motion.
\item We show that the learned representation supports both relative localization and a lightweight extension to few-shot absolute localization with sparse anchors.
\end{itemize}

% The remainder of this paper is organized as follows. Sec.~\ref{ch:problem_formulation} defines the problem setting. Sec.~\ref{ch:method} presents the proposed model. Sec.~\ref{ch:experiments} reports experimental results, and Sec.~\ref{ch:conclusion} concludes the paper.
\section{Problem Formulation}
\label{ch:problem_formulation}

We consider a 2D indoor field with $n$ WiFi access points. A moving device collects two synchronized traces along a trajectory of length $m$: an \emph{f-trace} $F=(f_1, \ldots, f_m)$ and a \emph{d-trace} $D=(d_1, \ldots, d_m)$. Each fingerprint $f_i=(s_1, \ldots, s_n)$ records RSSI values from nearby APs, while each displacement vector $d_i$ represents the stepwise motion between two adjacent sampling points. More specifically, if the trajectory positions are $(p_1, p_2, \ldots, p_m)$, then $d_i \approx \overrightarrow{p_{i-1}p_i}$ for $i=2,\ldots,m$, and we set $d_1=\vec{0}$ as the initial placeholder. We regard $D$ as a \emph{weak label} of $F$ because it contains only relative motions rather than absolute coordinates.

Given a crowdsourced dataset $\mathcal{D}$ of paired traces $(F,D)$, our goal is to learn a model that takes two f-traces $F_A$ and $F_B$ as input and predicts the displacement from the endpoint of $F_A$ to the endpoint of $F_B$. Formally, if the endpoints of the two traces are $p_A^m$ and $p_B^m$, respectively, then the target is
\begin{equation}
\Delta_{AB} = \overrightarrow{p_A^m p_B^m} = p_B^m - p_A^m.
\end{equation}
That is, we aim to learn a function
\begin{equation}
\mathrm{RelLoc}(F_A, F_B) \approx \Delta_{AB}.
\end{equation}
As illustrated conceptually in Fig.~\ref{fig:problem_definition}, this differs from conventional absolute localization, which first estimates two global coordinates and then subtracts them. Our formulation directly targets the relative displacement and is therefore better aligned with applications in which one mobile entity needs to locate another.

The key challenge is to make WiFi fingerprints \emph{computable} in the displacement domain. Fingerprints are ambiguous and noisy, whereas physical displacements are additive. The corresponding method design is presented in the next section.

\section{Method}
\label{ch:method}

\begin{figure}[t]
    \centering
    \includegraphics[width=0.4\textwidth]{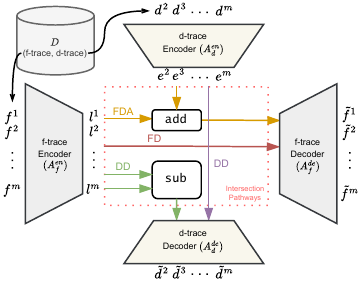}
    \caption{Training architecture of the proposed Intersection Pathway (IP), which takes paired \emph{f-trace} and \emph{d-trace} as input.}
    \label{fig:arch_train}
\end{figure}

\begin{figure}[t]
    \centering
    \includegraphics[width=0.4\textwidth]{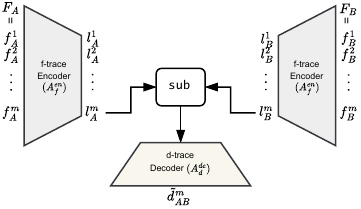}
    \caption{Inference architecture of the proposed Intersection Pathway (IP), which takes two \emph{f-traces} $(F_A, F_B)$ and predicts their relative displacement $\tilde{d}^m_{AB}$.}
    \label{fig:arch_infer}
\end{figure}

Figs.~\ref{fig:arch_train} and \ref{fig:arch_infer} illustrate the proposed framework. At a high level, our design consists of two autoencoders that share one latent space: one processes \emph{f-traces} and the other processes \emph{d-traces}. On top of these two branches, we construct direct pathways for reconstruction and cross-modal pathways for interaction between fingerprints and motions. This design allows local motion relations to regularize fingerprint relations, so that latent-space addition and subtraction become meaningful for relative-position inference. The autoencoders are therefore not introduced merely for exact input reconstruction; rather, they are used to learn a shared latent space in which fingerprint context can be regularized and displacement composition can be preserved. During training, the model takes a paired input $(F,D)$ from $\mathcal{D}$. During inference, it takes two \emph{f-traces} $F_A$ and $F_B$, and directly predicts the endpoint displacement from $A$ to $B$ by decoding the difference between their endpoint latent codes through the \emph{d-trace} branch.

\textbf{\emph{F-trace} autoencoder:}
An \emph{f-trace} is a short sequence of WiFi fingerprints, so similar fingerprints may appear at different positions and a single sample may fluctuate over time. To handle this ambiguity, we encode each \emph{f-trace} with a transformer-based autoencoder. The \emph{f-trace} encoder maps $F=(f_1,\ldots,f_m)$ to latent codes
\begin{equation}
A_f^{en}(F)=(l_1,\ldots,l_m),
\end{equation}
where each code captures both the current fingerprint and its sequential context. The corresponding decoder reconstructs the input trace as
\begin{equation}
A_f^{de}(l_1,\ldots,l_m)=(\tilde f_1,\ldots,\tilde f_m).
\end{equation}
This sequential modeling is important because the meaning of one fingerprint becomes clearer when its upstream and downstream context is also considered. The reconstruction branch is used to stabilize these context-aware embeddings under noisy RSSI observations, rather than to assume that unseen traces must be reconstructed perfectly.

\textbf{\emph{D-trace} autoencoder:}
Each displacement vector $d_i$ is mapped to a latent code $e_i$ by a linear autoencoder. The \emph{d-trace} encoder maps $D=(d_1,\ldots,d_m)$ to
\begin{equation}
A_d^{en}(D)=(e_1,\ldots,e_m),
\end{equation}
and the decoder maps the latent displacement codes back to reconstructed displacements,
\begin{equation}
A_d^{de}(e_1,\ldots,e_m)=(\tilde d_1,\ldots,\tilde d_m).
\end{equation}
We deliberately keep this branch linear so that physical additivity is preserved in the latent space. The latent dimensionality of $e_i$ is matched to that of $l_i$, enabling the two modalities to interact directly. Here the autoencoder serves to place displacement vectors into the shared latent space while preserving their additive structure, rather than merely reconstructing them. This linearity is essential because our inference relies on latent-space addition and subtraction to compose or compare displacements.

\textbf{Intersection Pathway:}
To align the two modalities, we use four training pathways based on the above encoders and decoders. The four pathways are defined as follows.

\emph{Fingerprint-Direct (FD):} this pathway reconstructs the input \emph{f-trace} and is trained with
\begin{equation}
L_{FD}=\sum_{i=1}^{m}|f_i-\tilde f_i|.
\end{equation}

\emph{Displacement-Direct (DD):} this pathway reconstructs the input \emph{d-trace} and is trained with
\begin{equation}
L_{DD}=\sum_{i=2}^{m}|d_i-\tilde d_i|.
\end{equation}

\emph{Fingerprint-Displacement-Add (FDA):} this pathway injects local motion into the fingerprint latent space. Here, $l_{i-1}$ is the latent code of fingerprint $f_{i-1}$, $e_i$ is the latent code of displacement $d_i$, $\hat l_i$ is the estimated latent code of the next fingerprint, and $\hat f_i$ is its decoded reconstruction. Intuitively, starting from the latent code of the previous fingerprint, we add the latent displacement corresponding to the motion from $p_{i-1}$ to $p_i$ and expect the result to match the latent code of the next fingerprint. Formally,
\begin{equation}
\hat l_i=
\begin{cases}
    l_i, & i=1,\\
    l_{i-1}+e_i, & i=2,\ldots,m,
\end{cases}
\end{equation}
and the resulting latent sequence is decoded as
\begin{equation}
(\hat f_1,\ldots,\hat f_m)=A_f^{de}(\hat l_1,\ldots,\hat l_m).
\end{equation}
This pathway encourages the model to make latent addition correspond to physical motion transitions in the fingerprint domain. The loss is
\begin{equation}
L_{FDA}=\sum_{i=1}^{m}|f_i-\hat f_i|.
\end{equation}

\emph{Fingerprint-Fingerprint-Subtract (FFS):} this pathway performs the reverse interaction. Here, $l_i-l_{i-1}$ represents the latent difference between two adjacent fingerprints, $\hat e_i$ is the estimated latent code of displacement $d_i$, and $\hat d_i$ is the decoded displacement in the physical space. It estimates local displacement from two adjacent fingerprint latents and requires the decoded result to match the stepwise motion label. Formally,
\begin{equation}
\hat e_i=
\begin{cases}
    \vec{0}, & i=1,\\
    l_i-l_{i-1}, & i=2,\ldots,m,
\end{cases}
\end{equation}
and the estimated displacement is decoded as
\begin{equation}
(\hat d_1,\ldots,\hat d_m)=A_d^{de}(\hat e_1,\ldots,\hat e_m).
\end{equation}
This pathway encourages latent subtraction between adjacent fingerprints to correspond to physical displacement. The loss is
\begin{equation}
L_{FFS}=\sum_{i=2}^{m}|d_i-\hat d_i|.
\end{equation}

Training proceeds in three stages:
\begin{equation}
\begin{aligned}
L_{phase1} &= L_{FD}, \\
L_{phase2} &= L_{FD}+L_{FDA}, \\
L_{phase3} &= L_{FD}+L_{DD}+L_{FFS}.
\end{aligned}
\end{equation}
In the first stage, we use only $L_{FD}$ so that the model can learn robust latent representations of \emph{f-traces} from sequential WiFi fingerprints, which is particularly important for resolving fingerprint ambiguity. In the second stage, we introduce $L_{FDA}$ to inject local motion information into the fingerprint latent space, so that latent addition starts to reflect transitions between adjacent fingerprints. In the third stage, we further optimize $L_{DD}$ and $L_{FFS}$ so that physical displacements can be stably mapped to and from the latent space, and latent differences between adjacent fingerprints become aligned with displacement vectors.

At inference time as shown in Fig.~\ref{fig:arch_infer}, two \emph{f-traces} $F_A$ and $F_B$ are encoded into $(l_A^1,\ldots,l_A^m)$ and $(l_B^1,\ldots,l_B^m)$. The predicted relative displacement is
\begin{equation}
\hat{\Delta}_{AB}=\tilde{d}^m_{AB}=A_d^{de}(l_B^m-l_A^m),
\end{equation}
where $A_d^{de}$ is the \emph{d-trace} decoder. Intuitively, the model translates endpoint differences in the fingerprint latent space into displacement vectors in the physical space. The linear \emph{d-trace} branch is critical here because it preserves the additivity needed to generalize from short stepwise supervision to longer relative displacements.
\section{Experiments}
\label{ch:experiments}

\subsection{Dataset}

\begin{figure}[t]
    \centering
    \includegraphics[width=0.45\textwidth]{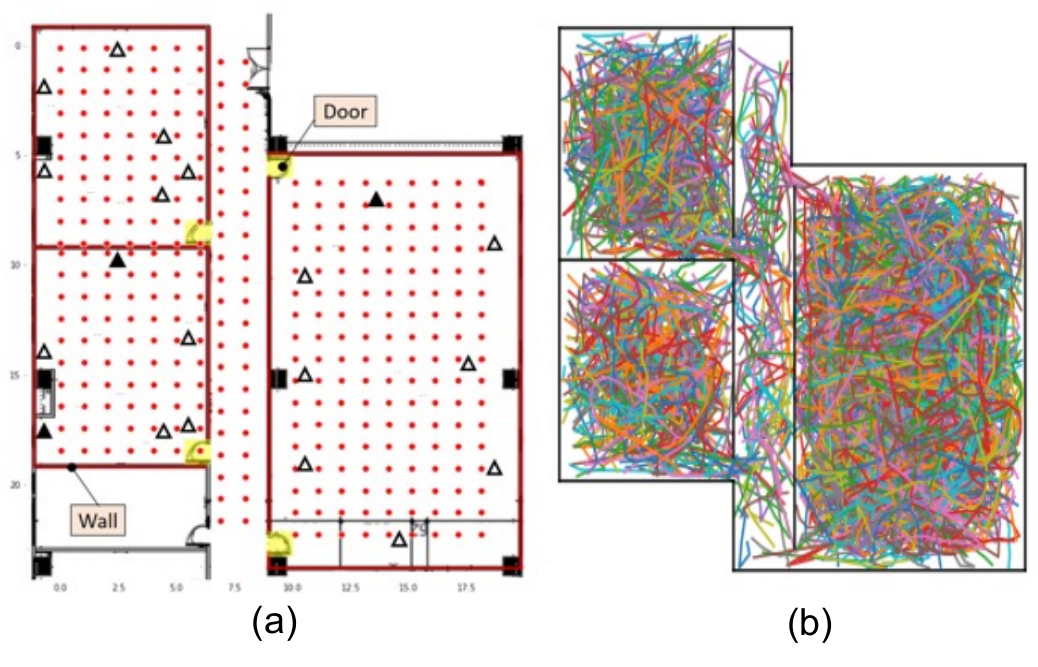}
    \caption{Dataset collection: (a) locations with manually collected WiFi fingerprints and AP placements, where solid and hollow triangles denote 2.4 GHz and dual-band (2.4+5 GHz) APs, respectively; and (b) synthesized trajectories in the target field.}
    \label{fig:data_map}
\end{figure}

Our experimental field is a concrete indoor space of roughly $20\times25$ meters with several rooms and a corridor, as illustrated in Fig.~\ref{fig:data_map}(a). It contains 20 WiFi APs and 337 labeled locations, with around 50 RSSI fingerprints collected at each location. Since a large real paired dataset of \emph{f-traces} and \emph{d-traces} is difficult to obtain, we build a near-real dataset for training and evaluation.

We collect (i) dense WiFi fingerprints in the experimental field and (ii) human walking trajectories in other fields, and synthesize a large dataset as follows.
\begin{enumerate}
\item
We repeatedly selected a trajectory from (ii) and cropped a length-$m$ sub-trajectory. These sub-trajectories formed a dataset $\mathcal{D}_{traj}$.
\item
We repeatedly sampled a trajectory $t \in \mathcal{D}_{traj}$ and a point $p$ in the experimental field. We shifted the starting point of $t$ to $p$ and randomly rotated it. If the trajectory does not hit any wall, we include it in a synthesized dataset $\mathcal{D}_{sy}$.
\item
From each trajectory in $\mathcal{D}_{sy}$, we derived a \emph{d-trace} and a \emph{f-trace}. For each point in the trajectory, if it fell in one of the labeled 337 locations, we sampled a WiFi fingerprint for it; otherwise, we synthesized one by Gaussian process regression (GPR)~\cite{williams1995gaussian}.
\end{enumerate}

Fig.~\ref{fig:data_map}(b) shows that trajectories in $\mathcal{D}_{sy}$ cover the target field well. The dataset is near-real rather than fully real-world: trajectory geometry comes from real motion traces, while RSSI fingerprints are queried from the target-field map after each transformed trajectory is placed in that field, with GPR used only for unlabeled points. To mimic deployment noise, we perturb both modalities. For any WiFi RSSI $f_u$, we inject noise by
\begin{equation}
    \hat{f_u} = f_u \times (1 + N(0, \sigma_f)),
\end{equation}
where $\sigma_f$ is a predefined standard deviation. For any displacement vector $(d_x, d_y)$, we convert it into a distance $r$ and an angle $\theta$, and perturb them by
\begin{equation}
    \begin{aligned}
        \hat{d_r} & = r \times (1 + N(0, \sigma_r)), \\
        \hat{d_\theta} & = \theta + N(0, \sigma_\theta).
    \end{aligned}
    \label{eq:displacement_r_noise}
\end{equation}

We then convert $(\hat{d_r}, \hat{d_\theta})$ back into a displacement vector $(\hat d_x, \hat d_y)$. Unless otherwise stated, we use 40K synthesized traces with $m=9$, set $(\sigma_f,\sigma_r,\sigma_\theta)=(0.05,0.2,0.15)$, and split the data into training and testing sets at a ratio of 8:2.

\subsection{Evaluation Metrics}

The relative localization problem has no widely adopted standard benchmark, so we use two complementary metrics.

\textbf{Displacement Error (DE):}
This metric measures the mean Euclidean error between the predicted relative displacement and the target relative displacement. Let
\begin{equation}
\hat{\Delta}_{AB} = A^{de}_{d}(l_B^m - l_A^m)
\end{equation}
be the predicted displacement from the endpoint of $F_A$ to that of $F_B$, and let $\Delta_{AB}$ denote the target displacement defined in Sec.~\ref{ch:problem_formulation}. We then define
\begin{equation}
DE = \frac{1}{N} \sum \left| \hat{\Delta}_{AB} - \Delta_{AB} \right|,
\end{equation}
where $N$ is the number of sampled pairs. Since our data provide only weak displacement supervision and displacement difficulty grows with distance, we further report
\begin{equation}
DE(k) = \frac{1}{N'} \sum_{|\Delta_{AB}| \le k} \left| \hat{\Delta}_{AB} - \Delta_{AB} \right|,
\end{equation}
where $N'$ is the number of sampled pairs whose target displacement length is at most $k$ meters. In particular, we report $DE(5)$, $DE(10)$, and $DE(all)$.

\textbf{Latent Code-Distance Ratio (LCDR):}
This metric evaluates whether latent-space distances scale consistently with physical displacements. Since sampled pairs may still have slightly different displacement lengths, we first normalize each latent distance by its target displacement length. Consider two sampled pairs $(F_A, F_B)$ and $(F_C, F_D)$ such that $|\Delta_{AB}| \approx |\Delta_{CD}|$. We define
\begin{equation}
    r_{AB} = \frac{|l_B^m - l_A^m|}{|\Delta_{AB}|}, \quad
    r_{CD} = \frac{|l_D^m - l_C^m|}{|\Delta_{CD}|}.
\end{equation}
If the learned latent space preserves displacement structure well, then pairs with similar physical displacements should also have similar normalized latent distances. We therefore define
\begin{equation}
LCDR_{A,B,C,D} = \frac{\min\{r_{AB}, r_{CD}\}}{\max\{r_{AB}, r_{CD}\}}.
\end{equation}
A value closer to 1 indicates better scaling consistency between latent-space distances and physical displacements. Over all sampled pairs, we compute
\begin{equation}
LCDR = \frac{1}{N} \sum LCDR_{A,B,C,D},
\end{equation}
and similarly define
\begin{equation}
LCDR(k) = \frac{1}{N'} \sum_{|\Delta_{AB}| \le k} LCDR_{A,B,C,D}.
\end{equation}

\subsection{Performance Evaluation}

Direct comparison with prior WiFi localization methods is not straightforward because most of them target absolute localization with strong labels \cite{du2020kf, localization-kuo, scrambling-kuo, boundary, wang2022research, ma2023few, 9918116}. In contrast, our task infers relative displacement from paired \emph{f-traces} under weak displacement supervision. We therefore focus on unified performance evaluation and controlled ablations, with few-shot absolute localization as a complementary reference.

\begin{table}[t]
    \vspace*{4pt}
    \centering
    \caption{Relative localization performance and key ablations.}
    \label{tb:main_relative}
    \begin{tabular}{lccc}
        \toprule
        Method & $DE(5)$ & $DE(10)$ & $DE(all)$ \\
        \midrule
        LSTM backbone & \textbf{0.300} & 0.885 & 1.818 \\
        DNN backbone & 0.575 & 1.034 & 1.961 \\
        w/o $A_d^{en}$ & 0.469 & 0.941 & 1.842 \\
        w/o $A_d^{en},A_f^{de}$ & 0.710 & 1.544 & 3.324 \\
        nonlinear $A_d$ & 1.505 & 2.707 & 4.672 \\
        \midrule
        Ours & 0.329 & \textbf{0.719} & \textbf{1.751} \\
        \bottomrule
    \end{tabular}
\end{table}

\begin{figure}[t]
    \centering
    \includegraphics[width=0.48\textwidth]{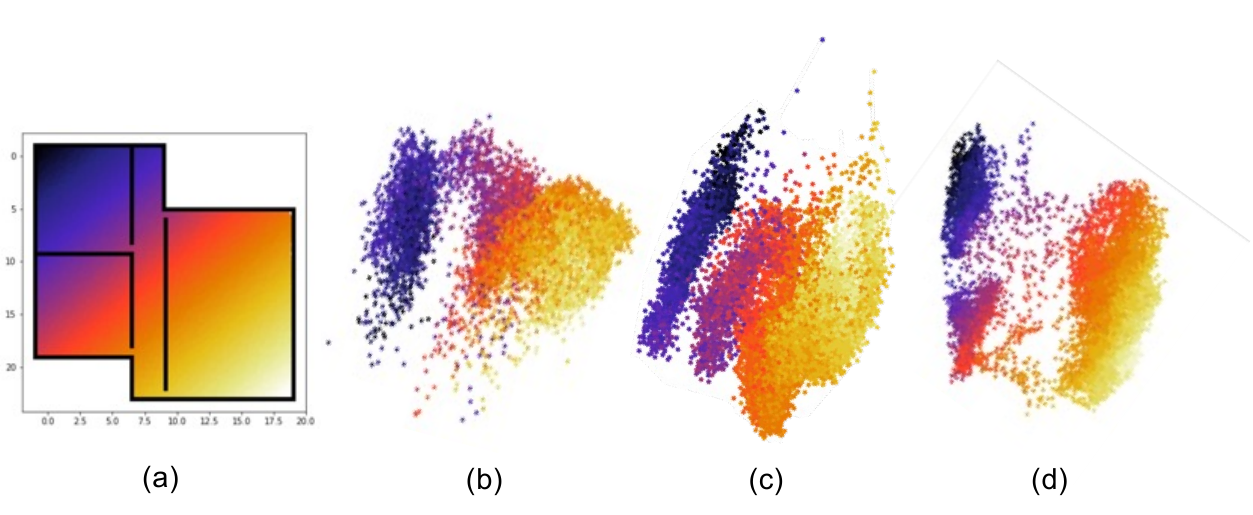}
    \caption{Coherence of latent codes with the physical map: (a) the physical map with color assignment by locations, and (b)--(d) PCA plots of endpoint latent codes after progressively longer stage-wise training.}
    \label{fig:epochs_test}
\end{figure}

Table~\ref{tb:main_relative} shows that our method achieves the best overall performance on $DE(10)$ and $DE(all)$, indicating that the shared latent space learned by IP captures longer-range spatial relations better than the alternatives. For the \emph{f-trace} backbone, the DNN variant performs poorly, while the LSTM backbone achieves the best $DE(5)$ but degrades more noticeably as the displacement range increases, suggesting that feed-forward or purely recurrent modeling is still limited for displacement composition over longer ranges.

The ablation variants further validate the proposed design. In ``w/o $A_d^{en}$'', the \emph{d-trace} encoder is removed, so displacement traces are no longer encoded into the shared latent space and the DD and FDA pathways are disabled accordingly. In ``w/o $A_d^{en},A_f^{de}$'', we further remove the \emph{f-trace} decoder, which also disables the FD pathway. The performance drops of these two variants show that both the cross-modal interaction and the reconstruction pathway contribute to stable relative-localization learning. Replacing the linear \emph{d-trace} branch by a nonlinear one (``nonlinear $A_d$'') leads to the largest degradation, confirming that linearity is important for preserving displacement composition in the latent space, which is exactly the property required by latent addition and subtraction during inference.

\begin{figure}[t]
    \centering
    \includegraphics[width=0.48\textwidth]{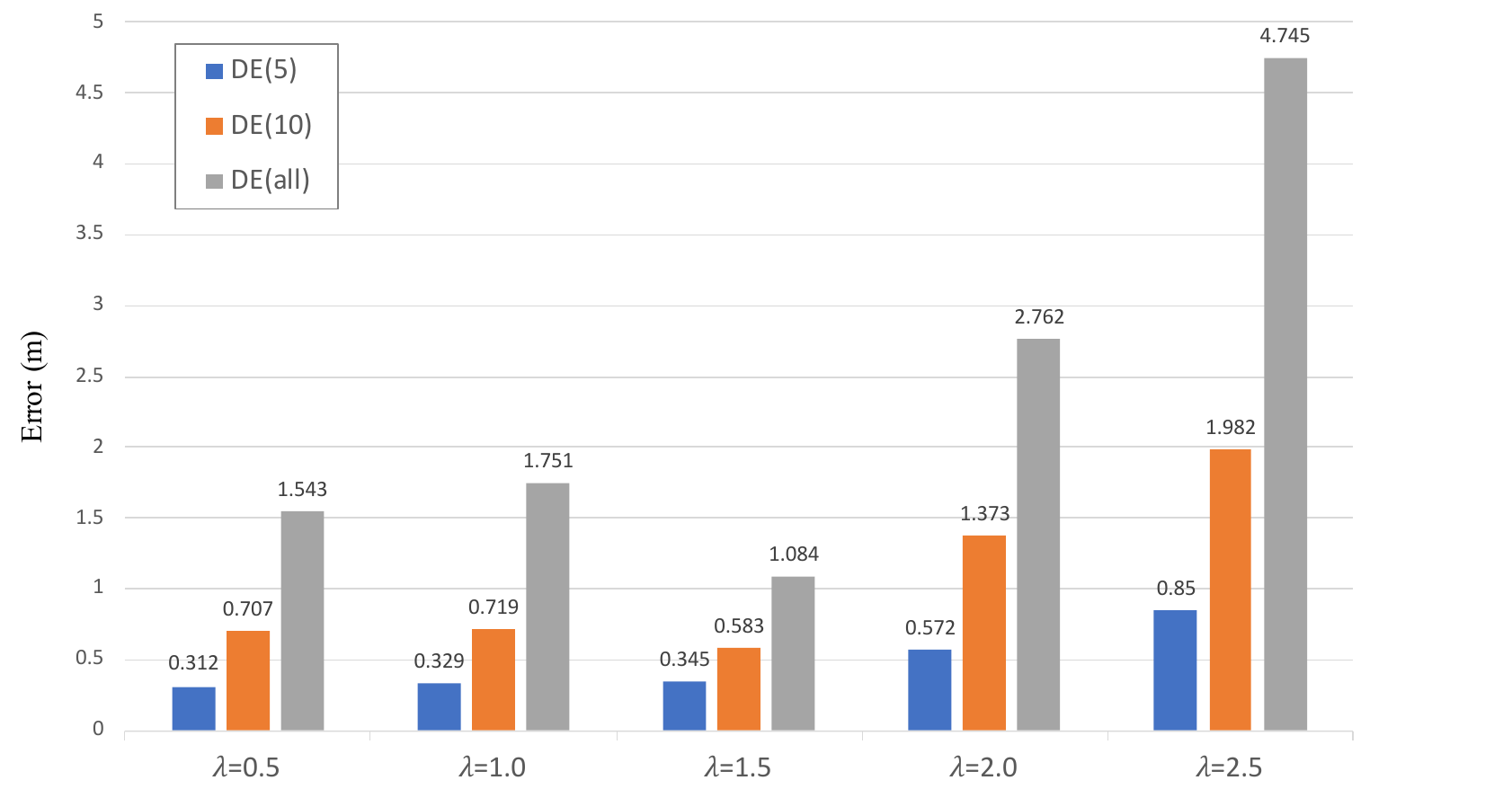}
    \caption{Impact of noise strength by varying $\lambda$.}
    \label{fig:lambda_experiment}
\end{figure}

\subsection{Training Dynamics and Sensitivity Analyses}

We further examine how the latent space evolves during training. Fig.~\ref{fig:epochs_test} shows the latent codes of the endpoints of 40K \emph{f-traces} after PCA projection. In the early stage, the model mainly captures local signal-spatial correlations. As stage-2 and stage-3 training continue, the latent codes become increasingly coherent with the physical layout, indicating that the proposed staged training gradually aligns fingerprint relations with displacement structure.

To evaluate robustness, we vary the noise strength for both \emph{f-traces} and \emph{d-traces} by scaling $(\sigma_f,\sigma_r,\sigma_\theta)$ with a factor $\lambda$. Fig.~\ref{fig:lambda_experiment} shows that the model remains stable when $\lambda \le 2.0$, but degrades sharply at $\lambda=2.5$. This suggests that the proposed framework can tolerate moderate sensing noise, while excessive perturbation weakens the learned fingerprint-displacement alignment.

Table~\ref{tb:size_dataset_experiment} further studies the impact of trace length and dataset size. Longer traces generally improve geometric consistency and long-range localization performance, although very short-range accuracy does not always monotonically improve. Increasing the dataset size consistently improves both DE and LCDR, confirming that the model benefits from broader spatial coverage and richer local transitions.

\begin{table}[t]
    \caption{Impact of trace length and dataset size.}
    \label{tb:size_dataset_experiment}
    \centering
    \renewcommand{\arraystretch}{1.1}
    \resizebox{\linewidth}{!}{
    \begin{tabular}{c|cccccc}
        \toprule
             & $DE(5)$ & $DE(10)$ & $DE(all)$ & $LCDR(5)$ & $LCDR(10)$ & $LCDR(all)$ \\ \hline
            $m=5$ & 0.574 & 1.138 & 2.272 & 0.686 & 0.595 & 0.542 \\
            $m=7$ & \textbf{0.322} & 0.813 & 2.077 & \underline{0.715} & 0.594 & 0.542 \\
            $m=9$ & \underline{0.329} & \underline{0.719} & \underline{1.751} & \textbf{0.744} & \underline{0.641} & \underline{0.592} \\
            $m=11$ & 0.446 & \textbf{0.559} & \textbf{1.037} & 0.690 & \textbf{0.642} & \textbf{0.638} \\ 
            \hline
            $|\mathcal{D}|=4K$ & 0.910 & 2.403 & 5.472 & \underline{0.636} & 0.523 & 0.446 \\
            $|\mathcal{D}|=12K$ & \underline{0.696} & \underline{1.272} & \underline{2.267} & 0.618 & \underline{0.554} & \underline{0.535} \\
            $|\mathcal{D}|=40K$ & \textbf{0.329} & \textbf{0.719} & \textbf{1.751} & \textbf{0.744} & \textbf{0.641} & \textbf{0.592} \\
        \bottomrule
    \end{tabular}
    }
\end{table}

\subsection{Few-shot Absolute Localization Extension}

\begin{table*}
    \vspace*{4pt}
    \centering
    \caption{Few-shot absolute localization under varying noise strengths ($\lambda$) and trace lengths ($m$).}
    \label{tb:varying_lambda}
    \begin{tabular}{c|cccccc|cccc}
        \toprule
             & no noise & $\lambda=0.5$ & $\lambda=1.0$ & $\lambda=1.5$ & $\lambda=2.0$ & $\lambda=2.5$ & $m=5$ & $m=7$ & $m=9$ & $m=11$ \\ \hline
            LSTM-loc & 0.129 & 0.144 & 0.146 & 0.170 & 0.181 & 0.196 & 0.171 & 0.159 & 0.146 & 0.138 \\
            BERT-loc & 0.039 & 0.060 & 0.064 & 0.097 & 0.119 & 0.135 & 0.070 & 0.070 & 0.064 & 0.062 \\
            ours & 0.140 & 0.312 & 0.329 & 0.345 & 0.572 & 0.850 & 0.574 & 0.322 & 0.329 & 0.446 \\
        \bottomrule
    \end{tabular}
\end{table*}

Although our model is trained for relative localization, it can be extended to absolute localization by using a sparse set of anchor traces with known endpoints. Given a query trace $F_B$, we compute its relative displacement to each anchor trace $F_A$, recover the endpoint of $F_B$ from the anchor endpoint plus the predicted displacement, and use the anchor producing the shortest predicted displacement as the final reference.

Table~\ref{tb:varying_lambda} summarizes the few-shot absolute-localization performance under varying noise strengths and trace lengths. LSTM-loc and BERT-loc are fully supervised baselines trained with strong labels for all traces, whereas our method uses weak labels for training and only sparse strong labels at inference time.

In the left half of Table~\ref{tb:varying_lambda}, the fully supervised baselines remain relatively stable as noise increases, whereas our method becomes more sensitive to stronger perturbation. In the right half, our method performs best at medium trace lengths, suggesting that very short traces provide insufficient context while overly long traces may accumulate more variation. Although our few-shot extension does not outperform fully supervised baselines, it achieves sub-meter accuracy while requiring far fewer strong labels.

\section{Conclusion}
\label{ch:conclusion}

We presented a weakly supervised framework for WiFi-based relative localization. Instead of predicting two absolute coordinates with dense strong labels, the proposed method directly estimates the displacement between two objects from their fingerprint traces. The core idea is Intersection Pathway, which aligns transformer-based WiFi embeddings and linear displacement embeddings in a shared latent space and makes latent addition/subtraction correspond to physical motion. Experiments show that this design yields strong relative-localization performance and that its key components, especially explicit cross-modal interaction and the linear \emph{d-trace} branch, are both important. We further showed that the learned model can be extended to few-shot absolute localization with sparse anchors. Future work may explore crowdsourced data under our framework.

% Our current evaluation is still limited to one indoor field and a near-real synthesized dataset built from real fingerprints and real trajectories. Future work includes validating the approach on larger fully real crowdsourced datasets, testing cross-building generalization, and extending it beyond 2D indoor environments.

\bibliographystyle{IEEEtran}
\bibliography{ref}

@ARTICLE{scrambling-kuo,
  author={Kuo, Sheng-Po and Tseng, Yu-Chee},
  journal={IEEE Transactions on Knowledge and Data Engineering}, 
  title={A Scrambling Method for Fingerprint Positioning Based on Temporal Diversity and Spatial Dependency}, 
  year={2008},
  volume={20},
  number={5},
  pages={678-684}}

@INPROCEEDINGS{localization-kuo,
  author={Kuo, Sheng-Po and Wu, Bing-Jhen and Peng, Wen-Chih and Tseng, Yu-Chee},
  booktitle={IEEE Int. Conf. on Mobile Adhoc and Sensor Systems}, 
  title={Cluster-Enhanced Techniques for Pattern-Matching Localization Systems}, 
  year={2007}}

@article{liu2020survey,
  title={Survey on WiFi-based indoor positioning techniques},
  author={Liu, Fen and Liu, Jing and Yin, Yuqing and Wang, Wenhan and Hu, Donghai and Chen, Pengpeng and Niu, Qiang},
  journal={IET communications},
  volume={14},
  number={9},
  pages={1372--1383},
  year={2020},
  publisher={Wiley Online Library}
}

@article{pasricha2020overview,
  title={Overview of indoor navigation techniques},
  author={Pasricha, Sudeep},
  journal={Position, Navigation, and Timing Technologies in the 21st Century: Integrated Satellite Navigation, Sensor Systems, and Civil Applications},
  volume={2},
  pages={1141--1170},
  year={2020},
  publisher={Wiley Online Library}
}

@ARTICLE{4343996,
  author={Liu, Hui and Darabi, Houshang and Banerjee, Pat and Liu, Jing},
  journal={IEEE Transactions on Systems, Man, and Cybernetics, Part C}, 
  title={Survey of Wireless Indoor Positioning Techniques and Systems}, 
  year={2007},
  volume={37},
  number={6},
  pages={1067-1080},
  doi={10.1109/TSMCC.2007.905750}
}

@article{shang2022overview,
  title={Overview of WiFi fingerprinting-based indoor positioning},
  author={Shang, Shuang and Wang, Lixing},
  journal={IET Communications},
  volume={16},
  number={7},
  pages={725--733},
  year={2022},
  publisher={Wiley Online Library}
}

@article{du2020kf,
  title={KF-KNN: Low-cost and high-accurate FM-based indoor localization model via fingerprint technology},
  author={Du, Canyang and Peng, Bao and Zhang, Zhaobo and Xue, Weicheng and Guan, Mingxiang},
  journal={IEEE Access},
  volume={8},
  pages={197523--197531},
  year={2020},
  publisher={IEEE}
}

@article{zhou2024vtil,
  title={VTIL: A multi-layer indoor location algorithm for RSSI images based on vision transformer},
  author={Zhou, Heng and Yang, Jingmin and Deng, Shanghui and Zhang, Wenjie},
  journal={Engineering Research Express},
  volume={6},
  number={1},
  pages={015069},
  year={2024},
  publisher={IOP Publishing}
}

@article{harle2013survey,
  title={A survey of indoor inertial positioning systems for pedestrians},
  author={Harle, Robert},
  journal={IEEE Communications Surveys \& Tutorials},
  volume={15},
  number={3},
  pages={1281--1293},
  year={2013},
  publisher={Ieee}
}

@article{wu2017automatic,
  title={Automatic radio map adaptation for indoor localization using smartphones},
  author={Wu, Chenshu and Yang, Zheng and Xiao, Chaowei},
  journal={IEEE Transactions on Mobile Computing},
  volume={17},
  number={3},
  pages={517--528},
  year={2017},
  publisher={IEEE}
}

@article{lashkari2018crowdsourcing,
  title={Crowdsourcing and sensing for indoor localization in IoT: A review},
  author={Lashkari, Bahareh and Rezazadeh, Javad and Farahbakhsh, Reza and Sandrasegaran, Kumbesan},
  journal={IEEE Sensors Journal},
  volume={19},
  number={7},
  pages={2408--2434},
  year={2018},
  publisher={IEEE}
}

@inproceedings{grosswindhager2019snaploc,
  title={SnapLoc: An ultra-fast UWB-based indoor localization system for an unlimited number of tags},
  author={Gro{\ss}windhager, Bernhard and Stocker, Michael and Rath, Michael and Boano, Carlo Alberto and R{\"o}mer, Kay},
  booktitle={Int'l Conf. on Information Processing in Sensor Networks},
  pages={61--72},
  year={2019}
}

@article{alhomayani2020deep,
  title={Deep learning methods for fingerprint-based indoor positioning: A review},
  author={Alhomayani, Fahad and Mahoor, Mohammad H},
  journal={Journal of Location Based Services},
  volume={14},
  number={3},
  pages={129--200},
  year={2020},
  publisher={Taylor \& Francis}
}

@article{wang2022research,
  title={Research on indoor 3D positioning algorithm based on wifi fingerprint},
  author={Wang, Lixing and Shang, Shuang and Wu, Zhenning},
  journal={Sensors},
  volume={23},
  number={1},
  pages={153},
  year={2022},
  publisher={MDPI}
}

@article{ma2023few,
  title={Few-Shot Learning for WiFi Fingerprinting Indoor Positioning},
  author={Ma, Zhenjie and Shi, Ke},
  journal={Sensors},
  volume={23},
  number={20},
  pages={8458},
  year={2023},
  publisher={MDPI}
}

@INPROCEEDINGS{9918116,
  author={Dong, Yinhuan and Arslan, Tughrul and Yang, Yunjie},
  booktitle={IEEE Int'l Conf. on Indoor Positioning and Indoor Navigation}, 
  title={An Encoded LSTM Network Model for WiFi-based Indoor Positioning}, 
  year={2022},
  volume={},
  number={},
  pages={1-6},
  doi={10.1109/IPIN54987.2022.9918116}
}

@article{zhao2020graphips,
  title={GraphIPS: Calibration-free and map-free indoor positioning using smartphone crowdsourced data},
  author={Zhao, Yonghao and Zhang, Zhixiang and Feng, Tianyi and Wong, Wai-Choong and Garg, Hari Krishna},
  journal={IEEE Internet of Things Journal},
  volume={8},
  number={1},
  pages={393--406},
  year={2020},
  publisher={IEEE}
}

@INPROCEEDINGS{piloc,
    author={Luo, Chengwen and Hong, Hande and Chan, Mun Choon},
    booktitle={Int'l Symp. on Information Processing in Sensor Networks}, 
    title={PiLoc: A self-calibrating participatory indoor localization system}, 
    year={2014},
    volume={},
    number={},
    pages={143-153},
    doi={10.1109/IPSN.2014.6846748}
}

@ARTICLE{9451560,
  author={Tan, Jiajie and Wu, Hang and Chow, Ka-Ho and Chan, S.-H. Gary},
  journal={IEEE Transactions on Mobile Computing}, 
  title={Implicit Multimodal Crowdsourcing for Joint RF and Geomagnetic Fingerprinting}, 
  year={2023},
  volume={22},
  number={2},
  pages={935-950}
}

@ARTICLE{boundary,
  author={Liu, Yu-Ting and Chen, Jen-Jee and Tseng, Yu-Chee and Li, Frank Y.},
  journal={IEEE Sensors Journal}, 
  title={An Auto-Encoder Multitask LSTM Model for Boundary Localization}, 
  year={2022},
  volume={22},
  number={11},
  pages={10940-10953}
}

@article{williams1995gaussian,
  title={Gaussian processes for regression},
  author={Williams, Christopher and Rasmussen, Carl},
  journal={Advances in neural information processing systems},
  volume={8},
  year={1995}
}

@inproceedings{chen2015blindnessavi,
  title={BlindNavi: A navigation app for the visually impaired smartphone user},
  author={Chen, Hsuan-Eng and Lin, Yi-Ying and Chen, Chien-Hsing and Wang, I-Fang},
  booktitle={ACM Conf. on Human Factors in Computing Systems},
  pages={19--24},
  year={2015}
}

\end{document}